\documentclass[twocolumn,amsmath,amssymb,aps,pre,showkeys,citesort]{revtex4-2}

\usepackage{cancel}
\usepackage{graphicx}
\usepackage{dcolumn}
\usepackage{bm}
\usepackage[caption=false]{subfig}
\usepackage{xcolor}

\definecolor{green}{RGB}{0,204,0}

\begin{document}

\title{Statistical properties of two-dimensional elastic turbulence}

\author{Himani Garg}
 \affiliation{Univ. Lille, ULR 7512 - Unité de Mécanique de Lille Joseph Boussinesq (UML), F-59000 Lille, France}

\author{Enrico Calzavarini}
 \affiliation{Univ. Lille, ULR 7512 - Unité de Mécanique de Lille Joseph Boussinesq (UML), F-59000 Lille, France}

\author{Stefano Berti}
 \affiliation{Univ. Lille, ULR 7512 - Unité de Mécanique de Lille Joseph Boussinesq (UML), F-59000 Lille, France}

\date{\today}

\begin{abstract}
We numerically investigate the spatial and temporal statistical properties of a dilute polymer solution in 
the elastic turbulence regime, \textit{i.e.}, in the chaotic flow state occurring at vanishing Reynolds and high Weissenberg numbers. 
\textcolor{black}{We aim at elucidating the relations between measurements of flow properties performed in the spatial domain with the ones taken in the temporal domain, which is a key point for the interpretation of experimental results on elastic turbulence and to discuss the validity of Taylor’s hypothesis.}
To this end, we carry out extensive direct numerical simulations of the two-dimensional Kolmogorov flow of an Oldroyd-B viscoelastic fluid.
Static point-like numerical probes are placed at different locations in the flow, particularly at the extrema of mean flow amplitude.
The results in the fully developed elastic turbulence regime reveal large velocity fluctuations, as compared to the mean flow, leading to a partial breakdown of Taylor's frozen-field hypothesis. 
While second-order statistics, probed by spectra 
and structure functions,  display consistent scaling behaviors  
in the spatial and temporal domains, the third-order statistics highlight robust differences. In particular the temporal analysis fails to capture the skewness of streamwise longitudinal  velocity increments.  
Finally, we assess both the degree of statistical inhomogeneity and isotropy of the flow turbulent fluctuations
as a function of scale. While the system is only weakly non-homogenous in the cross-stream direction, it 
is found to be highly anisotropic at all scales.
\end{abstract}

\maketitle

\section{\label{Sec:1}Introduction}

Viscoelastic fluids are known to be characterized by non-Newtonian behavior under appropriate conditions. In particular, 
dilute polymer solutions may display non-negligible elastic forces when the suspended polymeric chains occur to be sufficiently stretched by fluid velocity gradients. Remarkably, when the elasticity of the solution overcomes a critical value, such forces can trigger instabilities that can eventually lead to irregular turbulent-like flow, even in the absence of fluid inertia, 
namely in the limit of vanishing Reynolds number ($Re$). The latter dynamical regime is known as elastic turbulence (ET) \cite{GS00} 
and it has been experimentally observed in different flow configurations \cite{GS00,GS01,PMWA13,SACB17,sousa2018purely,Steinberg2021}. The emergence of this regime is characterized by the fast growth of the (largest) Lyapunov exponent of the flow as the polymer elasticity exceeds a system-dependent critical
value~\cite{Burghelea2004,Berti2008}. Besides, the kinetic-energy spatial and temporal spectra have a power-law behavior 
of exponent larger than $3$, in absolute value, which indicates the presence of a multiplicity of active scales in the flow. The spatial spectrum, however, is steeper than for Newtonian (three-dimensional) turbulence, {\it i.e.}, velocity fluctuations are concentrated at small wavenumbers; hence the velocity field is smooth in space \cite{GS00,Burghelea2007,Steinberg2021}. 

Based on its similarity with turbulent fluid motion, elastic turbulence has been proposed as an efficient system to enhance mixing in low-Reynolds-number flows \cite{GS01}. 
Moreover, it has been shown that it can increase heat transfer \cite{traore2015efficient,abed2016experimental}  
and promote emulsification \cite{poole2012emulsification}. 
Recently, it has also been argued that elastic turbulence flows play a significant role in the increased oil displacement obtained 
in industrial processes employing dilute polymer solutions to flood porous reservoir rocks \cite{MLHC16}.

\textcolor{black}{A compact and effective review of the main theoretical results on ET \cite{FL03} can be found in \cite{SACB17}, where several aspects of the agreement between the theory and experiments have been studied (see also \cite{Burghelea2007,Steinberg2021}).}
We have shown in previous work~\cite{Berti2008,BB10,Garg2018} that the main features of elastic turbulence can be numerically reproduced using a minimal flow model of a highly elastic polymer solution at low $Re$. 
Specifically, we considered the dynamics of Oldroyd-B model in a two-dimensional (2D) non-homogeneous, periodic, Kolmogorov flow setup.

Here, we perform a more thorough numerical study of the statistical properties of ET  in the same system. 
We focus on single- and multi-point statistics in both the temporal and spatial domains. We aim at elucidating the relations between measurements of flow properties  performed  in  the  spatial  domain  with  the  ones  taken  in  the  temporal  domain,  which is  a  key  point  for  the  interpretation  of  experimental  results  on  elastic  turbulence~\cite{Groisman2004,SACB17,Steinberg2021}. 
On the other hand, we aim to quantify the homogeneity and isotropy properties of 
the velocity fluctuations occurring in our flow. 
In particular, we 
examine the probability distribution function (PDF) of the fluctuating velocity and of its  derivatives, as well as the velocity power spectra, and the structure functions of velocity increments, in both the temporal and the spatial domain, at different flow locations.
We further address the validity of Taylor's frozen turbulence hypothesis as done in experiments~\cite{Burghelea2005}, in our numerical ET flows. 

This article is organized as follows. The model used to describe the dynamics of 
the viscoelastic fluid is introduced in Sec.~\ref{sec:2}. In Sec.~\ref{sec:3}, after briefly illustrating the main properties of the flow fields, we present the results 
about single- and two-point statistics both in the temporal and in the spatial domain. 
Conclusions are presented in Sec.~\ref{sec:5}.

\section{Model dynamics and simulations} \label{sec:2}

 \subsection{The 2D Kolmogorov flow for a viscoelastic fluid}
We consider the dynamics of a dilute polymer solution as described by the Oldroyd-B model~\cite{Bird1987}:
\begin{equation}
\partial_t \bm{u} + (\bm{u}\cdot\bm{\nabla}) \bm{u}  =  
- \frac{\bm{\nabla} p}{\rho_f} + \nu_s \Delta \bm{u} + {\frac{2\eta \nu_s}{\tau} \bm{\nabla} \cdot \bm{\sigma}} + \bm{f},
\label{eq:oldroyd_kf_u} 
\end{equation}
\begin{equation}
\partial_t \bm{\sigma} + (\bm{u} \cdot \bm{\nabla}) \bm{\sigma}  =  
(\bm{\nabla} \bm{u})^T \cdot \bm{\sigma} + \bm{\sigma} \cdot (\bm{\nabla} \bm{u}) - 2 \frac{\bm{\sigma}-\bm{1}}{\tau}.
\label{eq:oldroyd_kf_s}  
\end{equation}
In the above equations $\bm{u}=(u_x,u_y)$ is the incompressible velocity field, the symmetric positive definite matrix $\bm{\sigma}$ represents 
the normalized conformation tensor of polymer molecules \textcolor{black}{(normalized with the squared gyration 
radius)} 
and ${\bm 1}$ is the unit tensor corresponding to the equilibrium configuration 
of polymers attained in the absence of flow ($\bm{u}=0$). The trace $tr \left(\bm{\sigma}\right)$ gives the local polymer mean squared elongation while $\tau$ represents the largest typical polymer relaxation time. The fluid density is denoted ${\rho_f}$ and the total viscosity of the solution is 
$\nu = \nu_s(1 +\eta)$,  with $\nu_s$ the kinematic viscosity of the solvent and $\eta$ the zero-shear contribution of polymers (which is 
proportional to polymer concentration). The extra stress term ${\frac{2\eta \nu_s}{\tau} \bm{\nabla} \cdot \bm{\sigma}}$ accounts for 
elastic forces providing a feedback mechanism on the flow \textcolor{black}{\cite{balkovsky2001turbulence, FL03}}. 

In this study we are interested in a fluid velocity field characterized by a non-homogeneous mean flow and turbulent fluctuations 
generated by elastic stresses only. This is provided by the 2D periodic Kolmogorov flow. In the context of viscoelastic fluids, this flow  
has been previously adopted in~\cite{Berti2008,BB10,PGVG17} in order to provide a simple and effective model 
able to reproduce the basic phenomenology of elastic turbulence.     
Using the Kolmogorov forcing $\bm{f} = (F\cos(ky),0)$ in Eq. (\ref{eq:oldroyd_kf_u}), one has a 
laminar fixed point corresponding to the velocity field $\bm{u}^{(0)}=(U_0\cos(ky),0)$ and the conformation tensor components 
$\sigma^{(0)}_{11}=1+\frac{\tau^2 k^2 U_0^2}{2}\sin^2(ky)$, 
$\sigma^{(0)}_{12}=\sigma^{(0)}_{21}=-\frac{\tau k U_0}{2}\sin(k y)$, 
$\sigma^{(0)}_{22}=1$, with $F=  \nu k^2 U_0$ \cite{BCMPV05}. From these expressions, 
the characteristic length and velocity scales $L=1/k$ and $U_0$, respectively, can be identified.
As previously documented, the laminar flow becomes unstable \cite{BCMPV05} for sufficiently high values of elasticity, 
even in the absence of fluid inertia, and eventually displays features typical of turbulent flows \cite{Berti2008,BB10,Garg2018,PGVG17}. 
In the elastic turbulence regime, the mean velocity and conformation tensor fields keep similar simple sinusoidal functional 
forms but with different amplitudes. 
Denoting 
$U(y) = L_0^{-1}\int_0^{L_0} u_x(x,y)\ dx$ 
the mean longitudinal velocity, and $U$ its amplitude, where $L_0$ is the domain size (along the $x$-axis), we define the Reynolds number as $Re=UL/\nu$ and the Weissenberg number as $Wi=U\tau/L$.

\subsection{Numerical methods}
Equations (\ref{eq:oldroyd_kf_u}) and (\ref{eq:oldroyd_kf_s}) are integrated using a pseudo spectral method on a grid of side $L_0=2\pi$ 
with periodic boundary conditions at resolution $512^2$.  \textcolor{black}{We verified that the simulation results are independent of the mesh size.}
Integration of viscoelastic models is limited by instabilities associated with the loss of positiveness of the conformation 
tensor \cite{sureshkumar1995effect}. These instabilities are particularly relevant at high $Wi$ values and limit 
the possibility to numerically investigate the elastic turbulence regime by direct implementation of the equations of motion. 
For this reason we adopt an algorithm based on a Cholesky decomposition of the conformation matrix 
ensuring symmetry and positive definiteness \cite{vaithianathan2003numerical,Pinho2021}; to stabilize the code a small polymer diffusivity,
corresponding to a Schmidt number $Sc \approx 10^3 $, was also added. For Kolmogorov  and similar external forces that mix strain and vorticity, the effect of artificial polymer diffusivity may be less dramatic than in purely elongational flows~\cite{gupta2019effect,canossi2020elastic}. This approach allows us to reach sufficiently 
high elasticity; however it imposes some limitations in terms of resolution and computational time.
We fix $Re=1$ (using the value of $U_0$ to obtain an {\it a priori} estimate of it), which is smaller than the critical 
value $\sqrt{2}$ of the Newtonian case \cite{1961Newtonian}, and we 
choose a value of $Wi$ 
larger than the critical one $Wi_c \approx 10$ 
corresponding to the onset of purely elastic instabilities \cite{BB10}. 
The other  
parameters of the viscoelastic dynamics are $U_0=4$, $L=1/4$, $\nu_s=0.769$, $\eta=0.3$\textcolor{black}{, $\tau=2.5$}. 
The initial condition is obtained by adding a small random perturbation to the fixed point solution $\bm{u}^{(0)}$, $\bm{\sigma}^{(0)}$, and the system is evolved in time until a statistically steady state is reached. 
\textcolor{black}{To accelerate
the convergence, indeed quite slow, of the velocity statistics, the results are obtained by further averaging over an ensemble of $5$ independent data sets,
each of which was obtained from simulations begun with slightly different random initial conditions.}
In a recent work~\cite{Garg2018}, we have already shown that 
the resulting dynamics are only quite weakly dependent on $Wi$ in the fully developed elastic turbulence regime. 
Hence, in the following we will only consider a single Weissenberg number, namely $Wi = 24.9$~\cite{Garg2018}. 

\begin{figure}[]
\begin{center}
\includegraphics[width=0.48\textwidth]{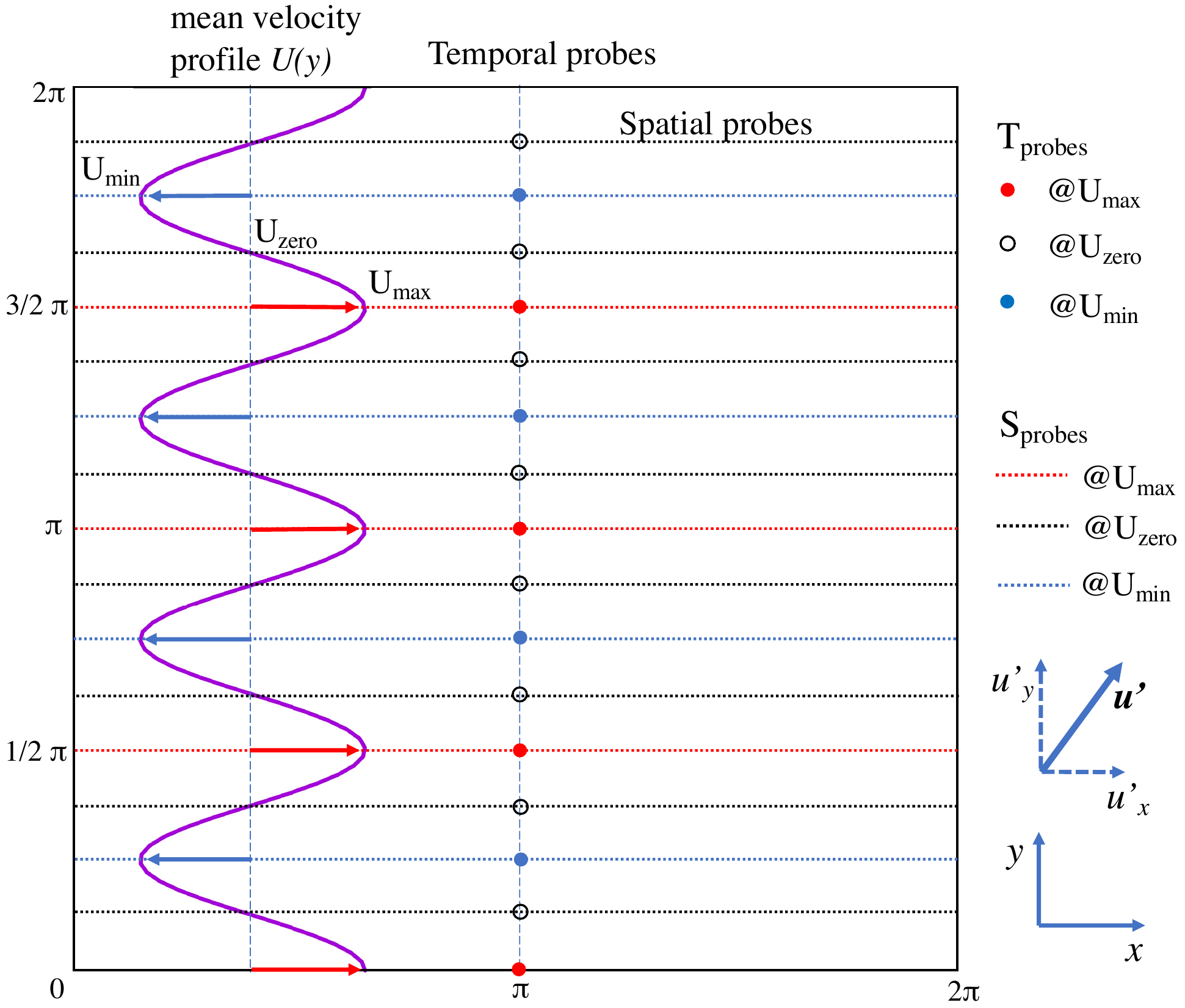}
\end{center}
\caption{
Positions of the different probes in the Kolmogorov flow setup. The large square box represents the spatial domain with size $[0,2\pi] \times [0,2\pi]$, while the solid violet line represents the mean velocity profile $U(y) = U \cos(k y)$, which has local maxima, minima and nodes denoted $U_{max},U_{min},U_{zero}$. The positions of the point-like temporal probes $T_{probes}$ are denoted by colored bullet symbols, while the positions of local spatial probes are indicated by the horizontal dashed lines.
}
\label{Fig:1a}
\end{figure}

In this study, we look at the flow in its statistically steady state from two different perspectives. The first is via spatial information obtained by instantaneous velocity fields stored during the simulations, 
the second is by 
analyzing time series taken at specific locations in the flow. While the first procedure 
is in principle preferable, as it 
may allow access to any relevant physical quantity, 
the second one is closer to the approach 
\textcolor{black}{sometimes} taken 
in experiments, in which the full dynamical field of interest is not always easily accessible \textcolor{black}{(but note that, through the Digital Particle Velocimetry technique, long time series of full spatial distributions of velocity were obtained for ET \cite{ SACB17, Burghelea2007, Burghelea2005}).}
To obtain the time series for the temporal analysis, we place static point-like temporal probes at different locations in the 
computational domain, in particular in the positions where the mean flow is maximum, minimum, and zero, denoted $@U_{max}$, $@U_{min}$ and $@U_{zero}$ (as shown in Fig.~\ref{Fig:1a}). 
We will then compare the measurements from these two viewpoints in a systematic way, 
highlighting the limitations intrinsically connected to probe-like single-point temporal data. 
\textcolor{black}{In fact, the availability of these two different data sets is interesting to assess the validity of Taylor’s hypothesis.}

\section{\label{sec:3}Results}

\subsection{The fluctuating velocity field}
\label{sec: fluct_flow}
\begin{figure*}[]
\begin{center}
\subfloat[]{
\includegraphics[width=0.3075\textwidth]{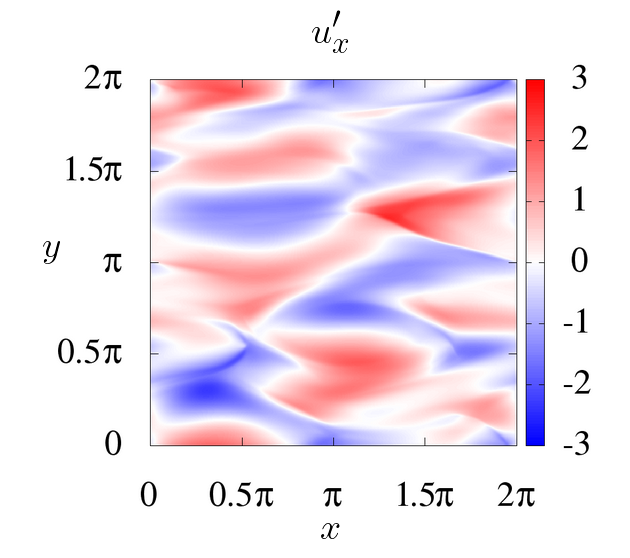}}
\subfloat[]{
\includegraphics[width=0.325\textwidth]{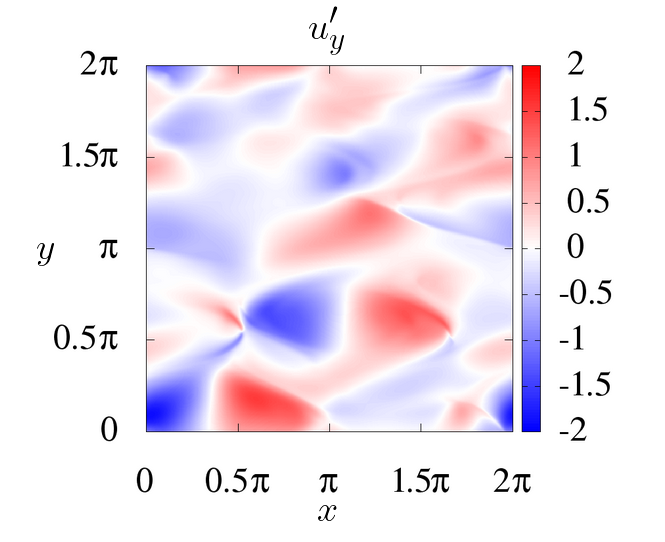}}
\subfloat[]{
\includegraphics[width=0.324\textwidth]{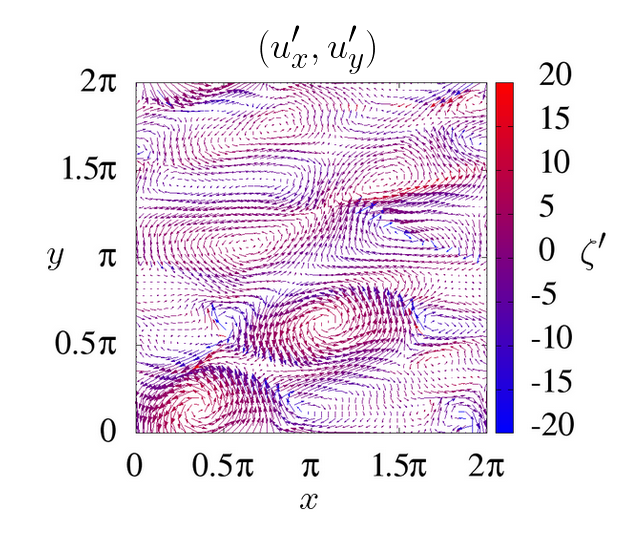}}

\subfloat[]{
\includegraphics[width=0.32\textwidth]{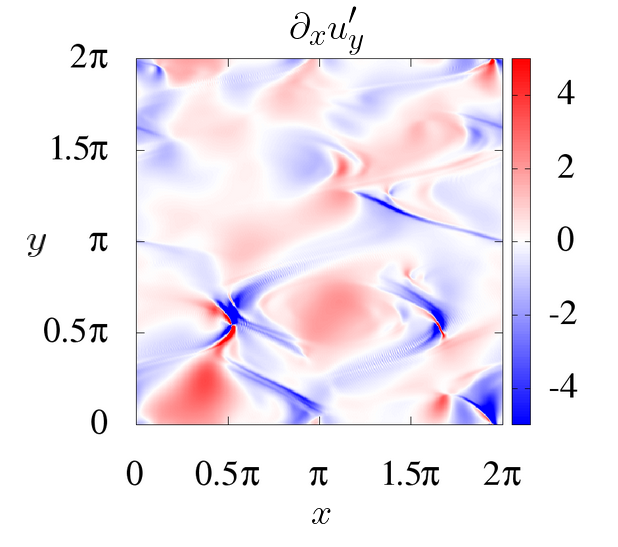}}
\subfloat[]{
\includegraphics[width=0.33\textwidth]{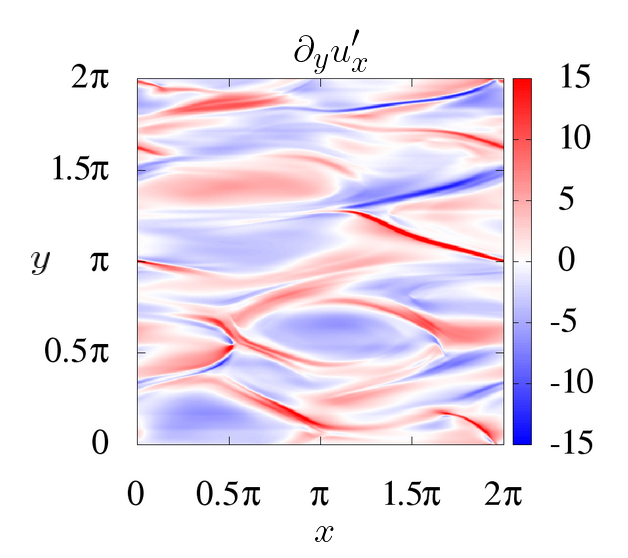}}
\subfloat[]{
\includegraphics[width=0.312\textwidth]{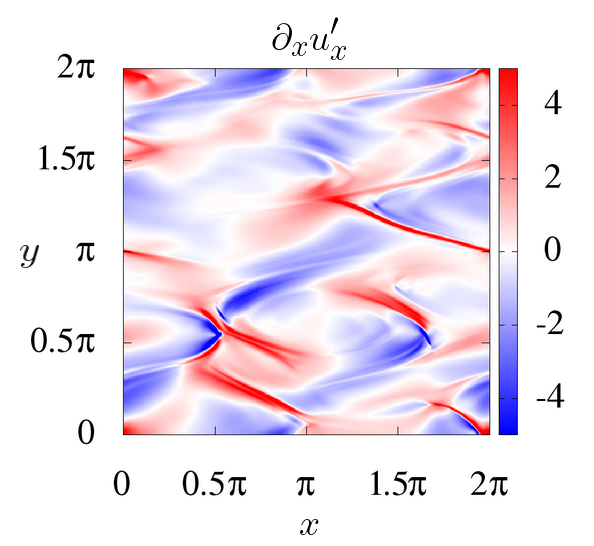}}

\caption{Instantaneous snapshots of $u'_x(x,y)$ (a),  $u'_y(x,y)$ (b), $\partial_x u'_y$ (d), $\partial_y u'_x$ (e), $\partial_x u'_x$ (f). 
Panel (c) shows the vector-field representation of the fluctuating velocity $\bm{u}'$ at the same instant of time as the other panels; here the color codes the intensity of the fluctuating vorticity $\zeta'$.}
\label{fig:distr_fields}
\end{center}
\end{figure*}

\begin{figure*}[]
\begin{center}
\subfloat[]{
\includegraphics[width=0.495\textwidth]{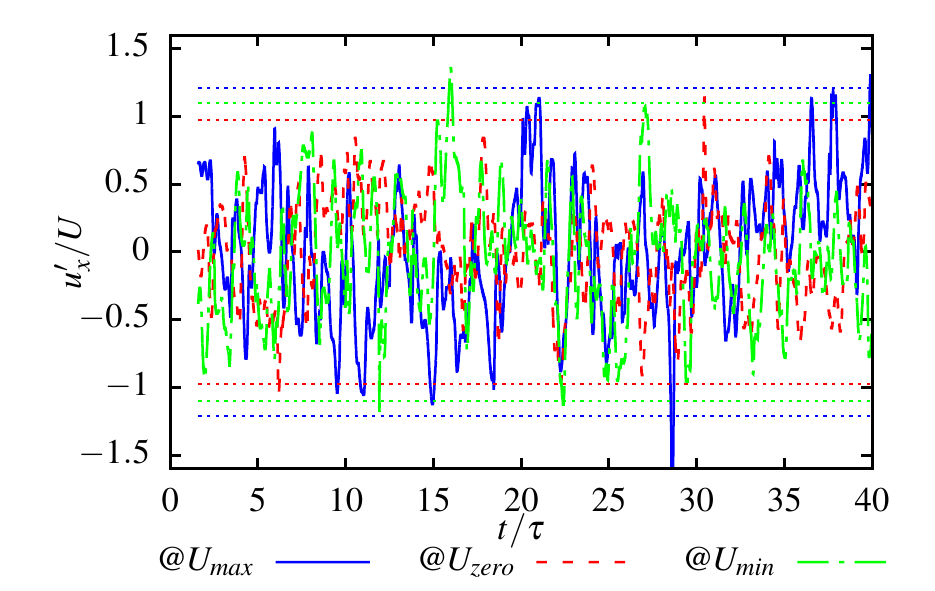}}
\subfloat[]{
\includegraphics[width=0.495\textwidth]{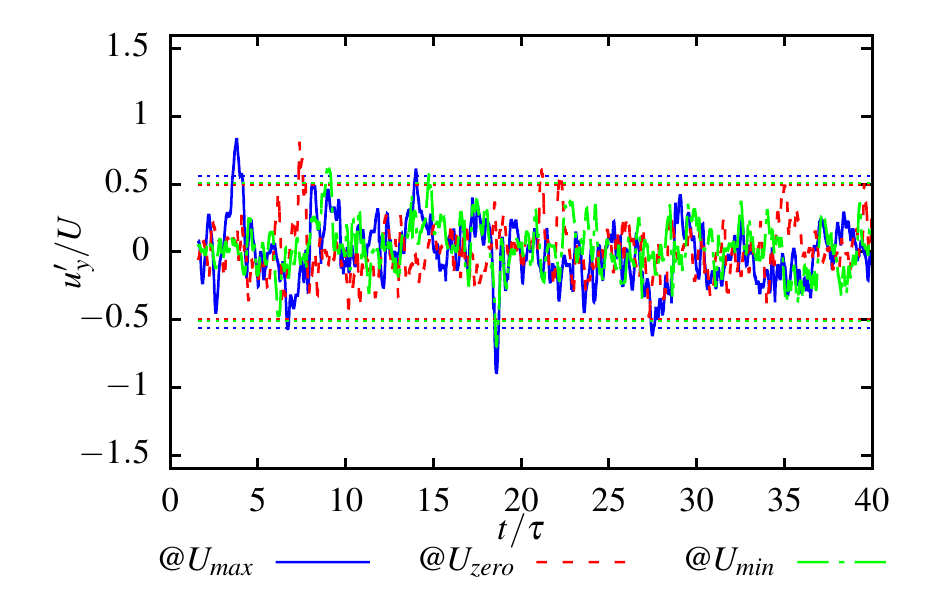}}
\caption{Time series of $u'_x$ (a) and $u'_y$ (b), normalized by the mean flow amplitude $U$, at $Wi=24.9$, as recorded by three different probes at the $U_{max}, U_{zero}, U_{min}$ locations.
\textcolor{black}{Horizontal dotted lines} indicate $\overline{u'_{x,y}} \pm \sigma_{u'_{x,y}}$, where the overbar denotes a temporal average and $\sigma_{u'_{x,y}}$ the standard deviation with respect to it. To ease visualization, only a short subset of the full data set (which reaches $t/\tau=800$) is shown.}
\label{fig: 2}
\end{center}
\end{figure*}

To look at the peculiar statistical features of the ET flow, it is convenient to consider the fluctuating component of the velocity field, defined as $\bm{u}'(x,y)=\bm{u}(x,y) - U \cos(k y) \hat{\bm{x}}$ (with $\hat{\bm{x}}$ the unit vector in the streamwise direction). Figure \ref{fig:distr_fields} displays an instantaneous snapshot of $\bm{u}'$ at $Wi = 24.9$. The figure shows separately the streamwise component $u_x'$ (panel a), the cross-stream component $u_y'$ (panel b), and the 
full vector field $\bm{u}'$ (panel c) 
colored according to the intensity 
of the instantaneous fluctuating vorticity field
$\zeta' = \partial_x u_y' - \partial_y u_x'$. Altogether, such a visualization allows us to easily appreciate the presence of secondary vortical structures 
emerging on top of the unidirectional mean background 
flow. 
These vortices appear to be stretched along the $y-$direction, probably as an effect of the mean flow. 
Their width along the $x$-direction carries the imprint of the characteristic length scale of the forcing. Overall, the secondary flow structures 
are observed to be organized in alternating stripes of counter-rotating eddies. Panels 
(d-f) in the same figure 
show the components of the velocity gradient tensor at 
the same instant of time. Note that only three independent components exist here due to 
incompressibility ($\partial_yu'_y = -\partial_xu'_x$). 
The spatial structure of the gradient fields reveals the existence of thin boundary regions between adjacent vortices. 
An example of the data obtained from point-like probes is provided in Fig.~\ref{fig: 2}. 
Specifically, here panels (a) and (b) respectively show the time series of the fluctuations of the streamwise velocity component, $u'_x$, and of the cross-stream one, $u'_y$, normalized by the mean flow intensity $U$, for three selected probes. In these plots, time is normalized by the polymer relaxation time $\tau$. 
Note that the data here reported correspond to a subset of the full time series, which extends to $800 \ \tau$. 
As one can clearly see, the temporal signals from different probes all share a common behavior, including that from the probe at zero-mean-flow location.
It is also worth remarking 
the different intensity of fluctuations 
in the $x$ and $y$ directions, which we will address more in detail in subsequent sections. 

\subsection{Temporal and spatial correlations of velocity fluctuations}
\label{sec:correl}
\begin{figure}[]
\begin{center}
\subfloat[]{
\includegraphics[width=1.0\linewidth]{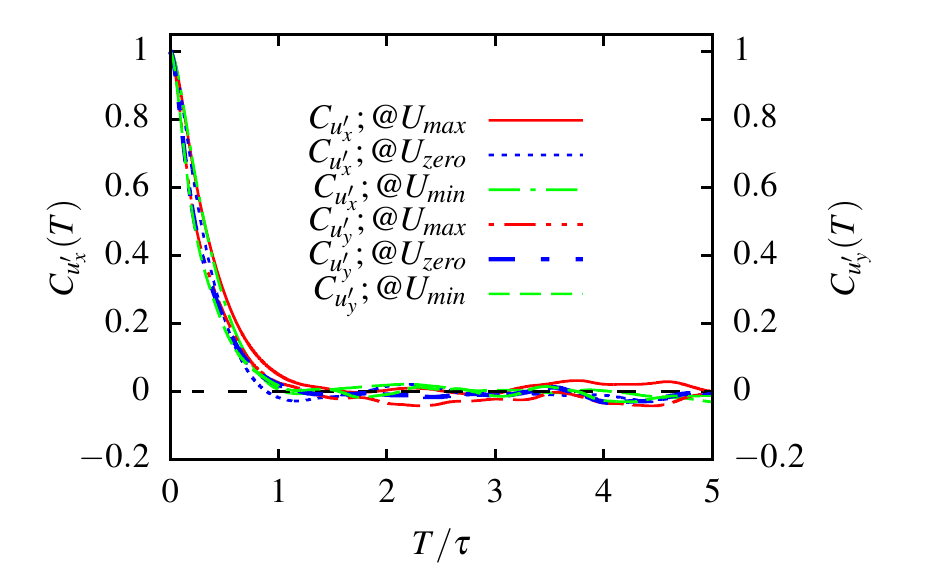}}

\subfloat[]{
\includegraphics[width=1.0\linewidth]{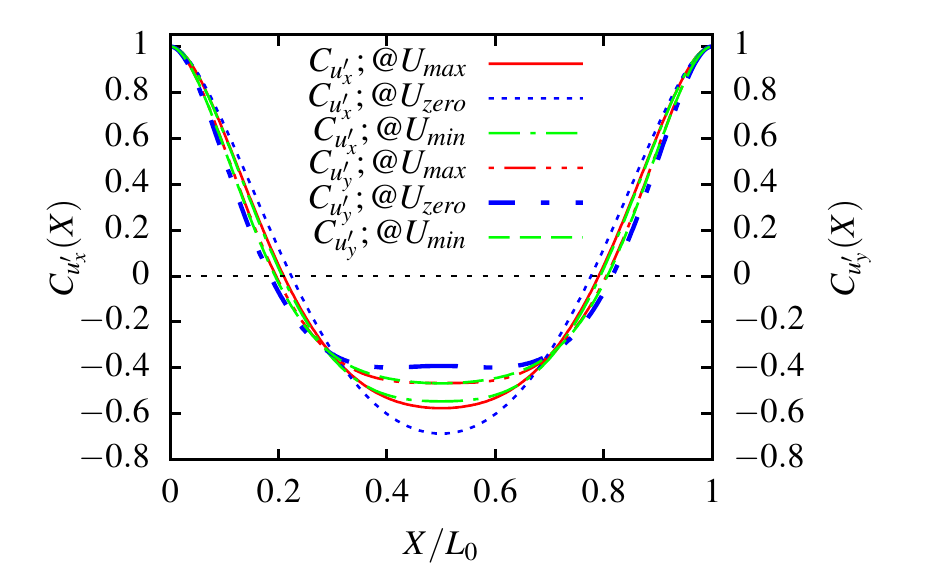}}
\caption{
Normalized temporal (a) and spatial (b) autocorrelation functions of $u'_x$ (left axes) and $u'_y$ (right axes). In both (a) and (b) data from probes where the mean flow is maximum, zero and minimum are shown. The Weissenberg number is $Wi=24.9$.}
\label{fig: Temporal autocorrelation length}
\end{center}
\end{figure}

Our measurements allow to quantitatively assess the correlations, in time and in space, of the fluctuating velocity field. Figure~\ref{fig: Temporal autocorrelation length}a shows the temporal correlation 
functions of the velocity fluctuation components, 
\begin{equation}
C_{u'_i}(T) = \frac{\langle u'_i(t+T)u'_i(t) \rangle_{t}}{\langle \left[u'_i(t)\right]^2 \rangle_{t}}, \; i=x, y.
\end{equation}

The corresponding correlation 
times, computed as $t_{u'_i} = \int_0^{T_{max}} C_{u'_i}(T)dT$ are reported in Table~\ref{table: temporal correlation time}, where $T_{max} = \mathcal{O}(10^2) \tau$ is the duration of the largest time increment. 
Irrespective of the position of the probe, we find that the correlation times of the two velocity components are comparable, and their values are of the order of the polymer relaxation time $\tau$. \textcolor{black}{This is in agreement with the theory by Lebedev and collaborators \cite{FL03} and experimental results \cite{SACB17}.}

The spatial correlation functions of each velocity component are defined as: 
\begin{equation}
 C_{u'_i}(X) = \frac{\langle u'_i(x+X)u'_i(x) \rangle_{x}}{\langle \left[u'_i(x)\right]^2 \rangle_{x}}, \; i=x, y,
 \label{Eq: spatial_correlation}
\end{equation}
where the brackets indicate an average over the $x$-direction and $X$ is the distance between two points in the same direction. Here we only focus on such a streamwise direction, which seems to us more interesting, as in the cross-stream one ($y$), the typical length scale is clearly constrained by the periodic structure of the mean flow and is therefore roughly equal to $1/k=1/4$ of the domain length.

We then determine the corresponding correlation lengths $l_{u'_i}$ as those for which the functions $C_{u'_i}(X)$ cross the zero line for the first time. The measured values are all quite close to $\approx 0.2 L_0$ (see Fig.~\ref{fig: Temporal autocorrelation length}b and Table~\ref{table: temporal correlation time}), which is close to the wavelength of the forcing, but slightly larger for $u'_x$, particularly where the mean flow vanishes, and the mean velocity gradient is largest.

\begin{table}[htpb]
\begin{center}
\begin{tabular}{|l|l|l|l|l|}
\hline
 & $ {t_{u'_x}}/{\tau}$ & $ {t_{u'_y}}/{\tau}$  & 
 $l_{u'_x}/L_0$ & $ {l_{u'_y}}/{L_0}$ \\\hline
$@U_{max}$   & 1.568  & 1.264 &  0.211 &  0.197   \\ \hline
$@U_{zero}$  & 0.885  & 1.286  &0.381 & 0.185            \\ \hline
$@U_{min}$   & 1.040  & 1.558  & 0.207 & 0.197         \\ \hline
\end{tabular}
\caption{
Correlation times and correlation lengths (in the streamwise direction) of fluctuations of both velocity components. The correlation times are measured as $t_{u'_i} = \int_0^{T_{max}}  C_{u'_i}(T)dT$, while the correlation lengths are deduced from the zero-crossing length of the corresponding autocorrelation functions $C_{u'_i}(X)$.}
\label{table: temporal correlation time}
\end{center}
\end{table}

\subsection{Analysis of probability density functions}

In this section, we focus on the probability density functions (PDF) of velocity fluctuations, as well as of the corresponding accelerations $\partial_t u'_i$ and of the spatial gradients $\partial_x u'_i$. Based on these analyses, we will also address the validity of Taylor's hypothesis for the present flow.

\subsubsection{PDF of velocity fluctuations}

An advantage of our numerical approach, with respect to experiments, is the possibility to use also information sampled at different locations in the flow. Hence, in addition to the data from the temporal probes ($T_{probes}$) introduced earlier, in the following we will also consider data extracted along lines in the streamwise direction (for fixed $y$ values corresponding to maximum, zero and minimum mean flow), as done for the correlation functions discussed in Sec.~\ref{sec:correl}. Aiming to compare the statistics from these data with those from single fixed-point measurements, we will refer to such data extractions as spatial probes ($S_{probes}$) (see Fig.~\ref{Fig:1a}).
The results already presented indicate a weak dependence of the statistical features of the flow from the position where the data are collected. For this reason, in order to enlarge the size of our samples, we will analyse together the data from probes in locations of maximum and minimum mean flow, and we will also consider global statistics, meaning obtained from all points in space (though for less instants of time).

\begin{figure}[htb]
\begin{center}
\subfloat[]{
\includegraphics[width=1\linewidth]{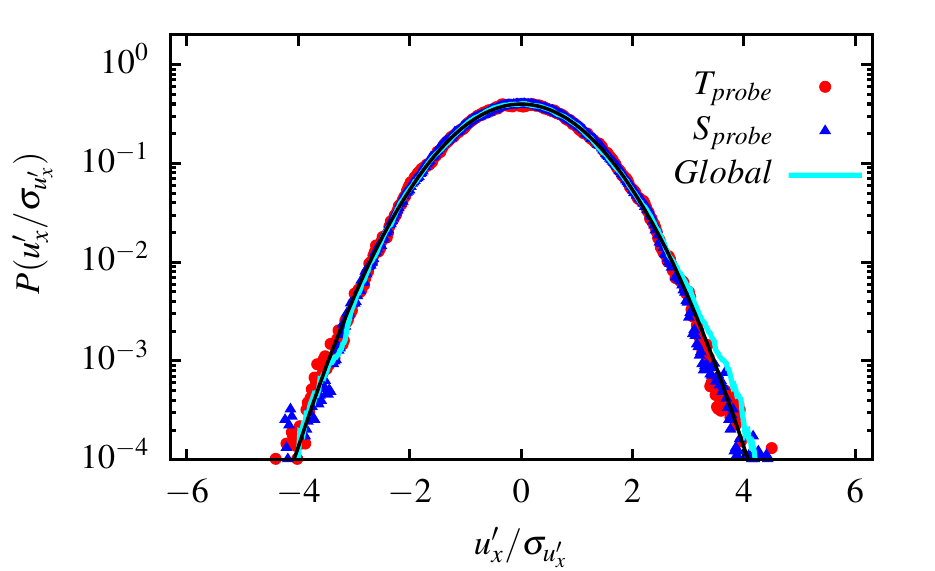}}

\subfloat[]{
\includegraphics[width=1\linewidth]{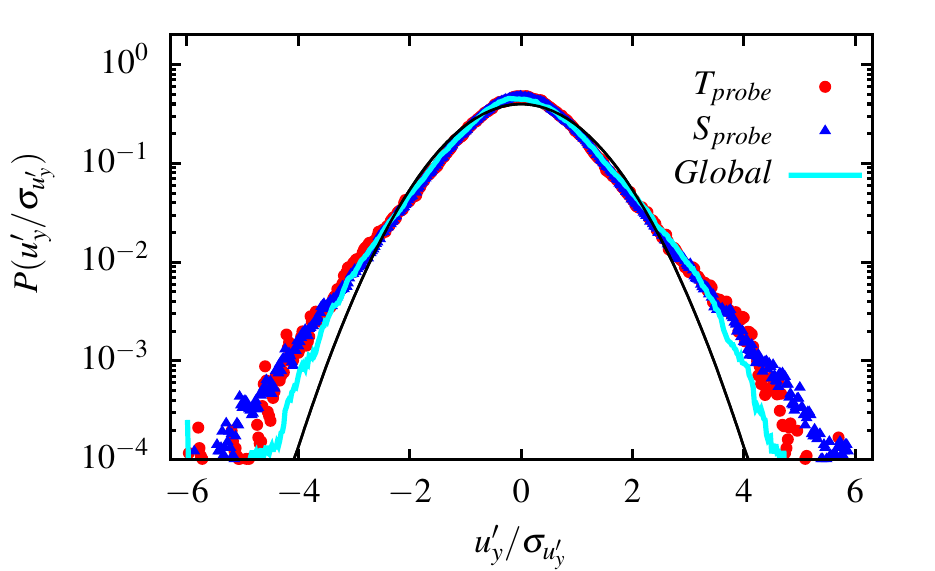}}

\caption{Probability distribution functions of (a) ${u'_x}/{\sigma_{u'_x}}$ and 
(b) ${u'_y}/{\sigma_{u'_y}}$ (with $\sigma_{u'_i}$ the rms values of $u'_i$ and $i=x,y$) for  $Wi=24.9$. 
Red \textcolor{black}{(circle)} and blue \textcolor{black}{(triangle)} points respectively correspond to the PDFs for temporal and spatial probes, using data collected where the mean flow amplitude is maximum and minimum. Aqua color points \textcolor{black}{(solid line)} stand for the PDFs from global spatial data from $1000$ instantaneous snapshots of velocity fields.}
\label{fig: pdf_velocity}
\end{center}
\end{figure}

In Fig.~\ref{fig: pdf_velocity} we present the
PDFs of both longitudinal and transversal 
velocity fluctuations for the data 
from temporal probes and  spatial 
ones (both locally and globally). 
Note that here the fluctuations are scaled by their root-mean-square (rms) value. 
In the present regime of fully developed elastic turbulence 
all the considered distributions of longitudinal velocity fluctuations 
are found to be well described by a Gaussian function 
with very little skewness  
(Fig.~\ref{fig: pdf_velocity}a).
In the transversal direction, the differently sampled PDFs are still close to each other, but we detect a slight deviation from Gaussianity, particularly for large fluctuations (Fig.~\ref{fig: pdf_velocity}b).
These curves are symmetric within our numerical accuracy. 
The values of different order moments 
(mean, $\langle ... \rangle$, standard deviation, $\sigma$, skewness, \textcolor{black}{$A$}, and kurtosis, $K$) of both $u'_x$ and $u'_y$, measured from temporal statistics, are reported in Table \ref{table: moments_velocity}. 

\begin{table*}[hbt]
\begin{center}
\begin{tabular}{|l|l|l|l|l|l|l|l|l|l|l|}
\hline
 {$T_{probes}$}& $\langle  u'_x \rangle$ & $\sigma_{u'_x}$ & \textcolor{black}{$A_{u'_x}$} & $K_{u'_x}$ & $\langle u'_y \rangle$ & $\sigma_{u'_y}$ & \textcolor{black}{$A_{u'_y}$} & $K_{u'_y}$ & $I_{u'_x}$ &  $I_{u'_y}$\\ \hline
$@U_{max}$   &2.469  & 1.186  & -0.059 &2.831 & -0.002  & 0.500  &0.042  & 4.235     &0.480 & 0.202\\ \hline
$@U_{zero}$  &-0.003  & 0.955  & 0.050 & 3.219 & -0.001 & 0.570 & 0.020  & 4.295       & &    \\ \hline
$@U_{min}$   &-2.460  & 1.170  &0.068  & 2.794  & -0.010 & 0.495  & -0.046   &  3.995  &0.476 & 0.201    \\ \hline \hline
Global   & -8.132e-08 & 1.034  & 0.020 &  2.789 & 4.142e-10 &  0.508 &  0.063  &   3.780 &  &     \\ \hline
\end{tabular}
\caption{
Moments of velocity fluctuations ($u'_x$ and $u'_y$): mean, $\langle ... \rangle$, standard deviation, $\sigma$, skewness, \textcolor{black}{$A$}, and kurtosis, $K$. In the last two columns we also report the  turbulence intensities for both velocity components, $I_{u'_x}$ and $I_{u'_y}$. All the values in this table are obtained from temporal probes at locations of maximum, zero and minimum mean flow. The last line corresponds to the global statistics. Note that the turbulence intensities are not defined in the absence of mean flow.}
\label{table: moments_velocity}
\end{center}
\end{table*}

\begin{figure}[htb]
\begin{center}
\subfloat[]{
\includegraphics[width=1\linewidth]{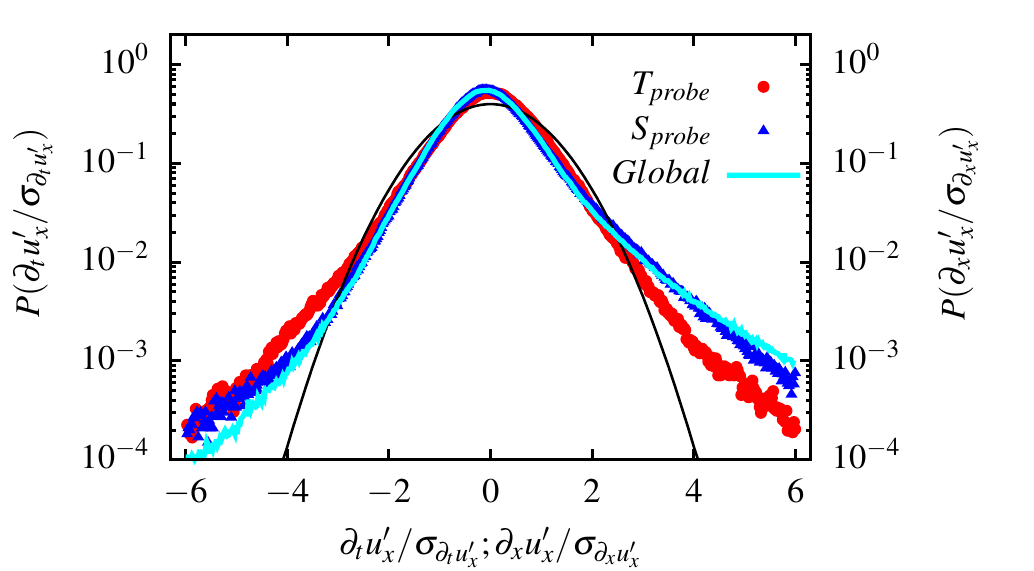}}

\subfloat[]{
\includegraphics[width=1\linewidth]{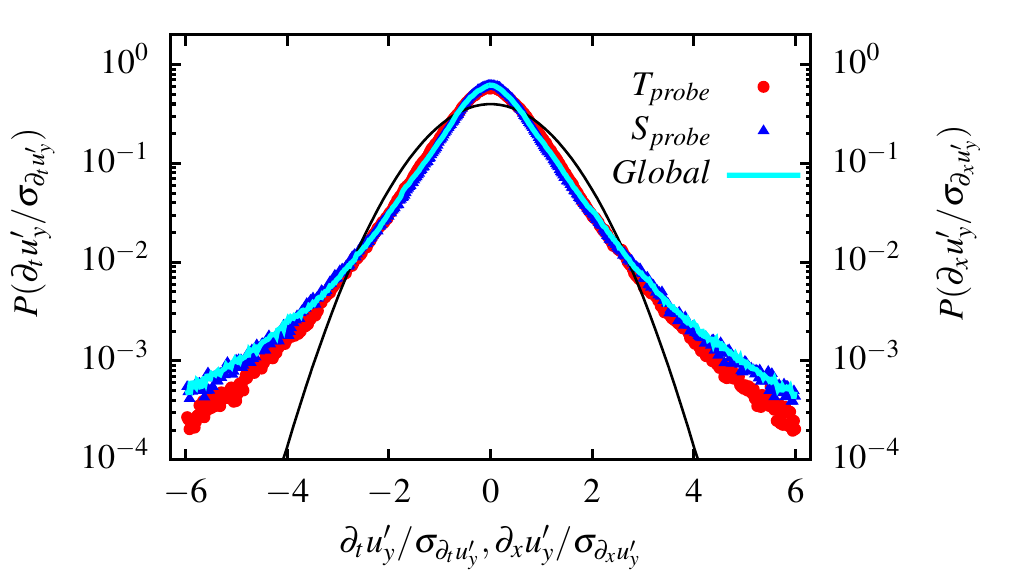}}
\caption{
Probability distribution functions of acceleration fluctuations ${\partial_tu'_i}/{\sigma_{\partial_tu'_i}}$ (red points \textcolor{black}{(circle)}, from temporal probes), and of velocity-gradient fluctuations ${\partial_x u'_i}/\sigma_{\partial_xu'_i}$, measured from both spatial probes (blue points \textcolor{black}{(triangle)}) and from global spatial data (aqua color points \textcolor{black}{(solid line)}), for $Wi=24.9$. All variables are scaled by their rms value $\sigma$. Panels (a) and (b) refer to the $x$ and $y$ components of velocity, respectively. The temporal and spatial probes are located where the mean flow is maximum and minimum, the global statistics make use of data from $1000$ instantaneous velocity fields.
}
\label{fig: pdf_accelerations}
\end{center}
\end{figure}

The probability distribution of Eulerian acceleration fluctuations $\partial_t u'_i = \frac{u'_i(t+\delta t)-u'_i(t)}{\delta t}$  
(with $i=x,y$) computed by finite differences with a  temporal increment $\delta t$, scaled by the rms value, is shown in Fig.~\ref{fig: pdf_accelerations} (red points \textcolor{black}{(circle)}, from temporal probes). 
For both $x$ and $y$ components, 
the PDFs deviate from the Gaussian distribution, especially for large fluctuations, providing evidence of intermittent behavior at small scales.
The situation of nearly Gaussian statistics of
velocities and essentially non-Gaussian, 
high-tail distributions of velocity derivatives, 
is similar to what occurs in high-$Re$ inertial turbulence~\cite{1995turbulence}.
This result is also consistent with the measurements \textcolor{black}{performed 
in} different ET geometrical setups \cite{Groisman2004,SACB17}\textcolor{black}{, but note that the PDFs of velocity derivatives (Fig.~\ref{fig: pdf_accelerations}) display slight deviations from the the exponentially tailed shape found in experiments.}
For completeness, as for the velocity statistics, the values of different order moments of acceleration statistics, measured from temporal probes, are reported in 
Table~\ref{table: moments_accelerations}. 

\begin{table*}[bth]
\begin{center}
\begin{tabular}{|l|l|l|l|l|l|l|l|l|}
\hline
$T_{probes}$ & $\langle \partial_t u'_x \rangle$ & $\sigma_{\partial_t u'_x}$ & \textcolor{black}{$A_{\partial_t u'_x}$} & $K_{\partial_t u'_x}$ & $\langle \partial_t u'_y \rangle$ & $\sigma_{\partial_t u'_y}$ & \textcolor{black}{$A_{\partial_t u'_y}$} & $K_{\partial_t u'_y}$  \\ \hline
$@U_{max}$   &0.022  & 4.922  & 0.637  & 15.925 & 0.000   & 2.869   & -0.032 & 12.633       \\ \hline
$@U_{zero}$  &-0.000  & 4.097  & -0.060 & 8.423& -0.000  & 3.200 & -0.042 & 10.753         \\ \hline
$@U_{min}$   & 0.021 & 4.933 & 0.704  & 16.684 & -0.001  & 2.910  & 0.049  & 12.736          \\ \hline
\end{tabular}
\caption{
Moments of acceleration fluctuations ($\partial_t u'_x$ and $\partial_t u'_y$): mean, $\langle ... \rangle$, standard deviation, $\sigma$, skewness, \textcolor{black}{$A$}, and kurtosis, $K$. 
All the values in this table are obtained from temporal probes at locations of maximum, zero and minimum mean flow.
}
\label{table: moments_accelerations}
\end{center}
\end{table*}

\subsubsection{The Taylor's frozen-field hypothesis and the PDF of velocity gradients} \label{sec: validation of TH}

We now address the validity of Taylor's hypothesis in the present 2D elastic turbulence model system. Taylor's hypothesis
states that for any quantity, $\phi$ , transported in a turbulent flow and measured at different times ($t$ and $t+T$) at a fixed location in space ($x$), the relation between temporal and spatial increments is the following:
\begin{equation}
 \phi(t+T) - \phi(t) =  \phi(x - U T) - \phi(x),
 \label{eq: Taylor hypothesis0}
\end{equation}
where $U$ is the mean velocity at the 
measurement location, 
$x$ is the spatial coordinate in the direction of the mean flow. This is equivalent to say that the conversion between temporal ($T$) and spatial ($X$) increments is $X = U \ T$ and it justifies dubbing it Taylor's frozen-field hypothesis (TH).
In the limit of vanishing increments $T$ Eq.~(\ref{eq: Taylor hypothesis0}) leads to 
\begin{equation}
 \partial_t \phi = -U\  \partial_x \phi.
 \label{eq: Taylor hypothesis}
\end{equation}
Note that the TH works also when the transported quantity is the turbulent velocity fluctuation itself, $u'_i$ or a function of it, such as the turbulent kinetic energy. 

Equations~(\ref{eq: Taylor hypothesis0})-(\ref{eq: Taylor hypothesis}) are commonly used, in experimental studies, to evaluate the streamwise spatial increments or derivative of $\phi$, given its finite increments or time derivative.
This is done by necessity 
since fully resolved 
flow-field measurements, in space, are notably difficult and require sophisticated experimental techniques. 
Besides the statistical steadiness of the flow the main requirement for the validity of TH
is the existence of a large local mean 
velocity, as compared to velocity
fluctuations \cite{Hinze1959}:
\begin{equation}
 \frac{u'}{U} << 1.
 \label{eq: TH1}
 \end{equation}
The extent to which TH is expected to hold, in general flow conditions, has been discussed  theoretically~\cite{taylor1938spectrum,Lin1953,Taylorhypothesis2}. However, this point needs, in principle, to be checked on a case-by-case basis in experiments and numerical simulations.

A large body of experimental works
addressed the subject, mainly in relation to two-dimensional and three-dimensional (3D) inertial turbulent flows~\cite{dahm1997experimental,2000_TH_ABELMONTE,TH_2D_PIVmeasurement,TH_2D_turbulentjet}.
In particular, detailed discussions have been devoted to the applicability of TH for flows with strong shear and high turbulence intensities \cite{Taylorhypothesis2,Cenedese_1991,dahm1997experimental,hutchins_marusic_2007,dennis_nickels_2008,TH_2D_turbulentjet,geng_2015}. 
Using accurate time-dependent measurements of the flow field, the degree of validity of TH 
was assessed in experiments of elastic turbulence, by examining the velocity coherence of pairs of points separated by both spatial and temporal intervals~\cite{Burghelea2005}.

To test the validity of the frozen-field hypothesis, one can either measure the velocity correlation functions or, in light of Eq.~(\ref{eq: Taylor hypothesis}), compare the accelerations $\partial_t u_i'$ with the velocity gradients $\partial_x u_i'$. 
These two approaches are complementary: the correlation-based method is conditioned by the spatial resolution of the velocity fields, whereas the second one requires a high temporal resolution. Here, we consider the second approach, choosing to compare the probability distribution functions of the temporal and spatial derivatives of velocity fluctuations.

To check whether the velocity fluctuations are transported without evolving dynamically, the essential meaning of TH, we compute the spatial gradients of each component of the velocity fluctuations along the streamwise direction of the flow, and compare them to the Eulerian accelerations previously calculated  (see Fig.~\ref{fig: pdf_accelerations}, where blue points \textcolor{black}{(triangle)} correspond to spatial probes and aqua color \textcolor{black}{(solid line)} ones to global statistics).
Table \ref{table: moments_spatial_gradients_global} 
provides a summary of the values measured for different order moments for both components of (longitudinal) spatial gradients. \textcolor{black}{The magnitude of the typical gradients at the present Weissenberg number, estimated from their standard deviations, is roughly of order $1/\tau$, in reasonable agreement  with theoretical predictions \cite{FL03} and independent measurements based on the Lagrangian Lyapunov exponent of the flow \cite{Berti2008} .}
\begin{table*}[hbt]
\begin{center}
\begin{tabular}{|l|l|l|l|l|l|l|l|l|}
\hline
$S_{probe}$ & $\langle \partial_x u'_x \rangle$ & $\sigma_{\partial_x u'_x}$ & \textcolor{black}{$A_{\partial_x u'_x}$} & $K_{\partial_x u'_x}$ & $\langle \partial_x u'_y \rangle$ & $\sigma_{\partial_x u'_y}$ & \textcolor{black}{$A_{\partial_x u'_y}$} & $K_{\partial_x u'_y}$  \\ \hline
 $@U_{max}$   &2.982e-10  & 1.292  & 0.909 & 8.032 & -6.535e-11  & 1.004  &0.048  & 6.001   \\ \hline
$@U_{zero}$  &-1.220e-10  & 1.220  & 0.976 & 7.127 & 6.138e-12 & 0.902 & 0.015  & 5.120      \\ \hline
$@U_{min}$   &-1.074e-10  & 1.234  & 0.923  & 7.234  & -7.582e-11 & 0.894  & -0.059   & 5.262      \\ \hline \hline
Global & -4.692e-12  & 1.240& 0.763 & 6.292 &-2.958e-12  & 0.965 & -0.036 & 6.086   \\ \hline
\end{tabular}
\caption{Moments of the longitudinal gradients of velocity fluctuations ($\partial_x u'_x$ and $\partial_x u'_y$): mean, $\langle ... \rangle$, standard deviation, $\sigma$, skewness, \textcolor{black}{$A$}, and kurtosis, $K$. 
All the values in this table are obtained from local spatial probes at locations of maximum, zero and minimum mean flow, global statistics and averaging over time.
}
\label{table: moments_spatial_gradients_global}
\end{center}
\end{table*}

Inspection of Fig.~\ref{fig: pdf_accelerations} reveals different statistical behaviors of temporal and spatial derivatives of the velocity fluctuations, at least for the flow component in the streamwise direction (Fig.~\ref{fig: pdf_accelerations}a). Indeed, while the distributions from both temporal and spatial probes exhibit high tails indicative of intermittency, the PDF of $\partial_x u'_x$ is found to be more positively skewed. Such a discrepancy is not in agreement with Eq.~(\ref{eq: Taylor hypothesis}) and, therefore, provides a first evidence of a partial breakdown of TH.

\begin{figure}[htb]
\begin{center}
\subfloat[]{
\includegraphics[width=1\linewidth]{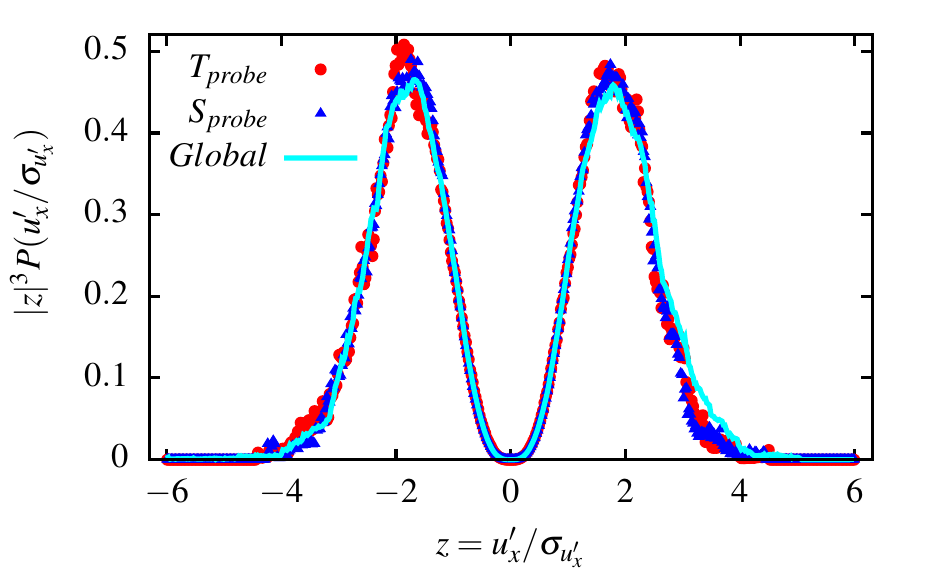}}

\subfloat[]{
\includegraphics[width=1\linewidth]{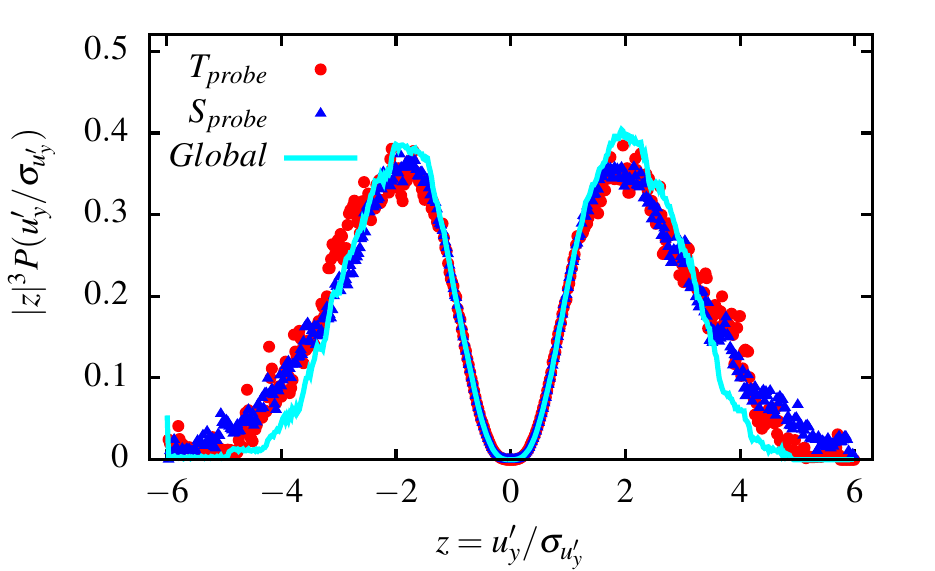}}
\caption{Third-order statistics $|z|^3P(z)$ for $z=u'_x/\sigma_{ u'_x}$ (a) and $z=u'_y/\sigma_{ u'_y}$ (b), for $Wi=24.9$. Red \textcolor{black}{(circle)} and blue \textcolor{black}{(triangle)} points respectively correspond to data from temporal and spatial probes, located where the mean flow is maximum and minimum; aqua color points \textcolor{black}{(solid line)} correspond to global spatial data. All spatial statistics are further averaged over time using $1000$ instantaneous velocity fields.
}
\label{fig: pdf moments of order 3}
\end{center}
\end{figure}

\begin{figure}[htb]
\begin{center}
\subfloat[]{
\includegraphics[width=1.1\linewidth]{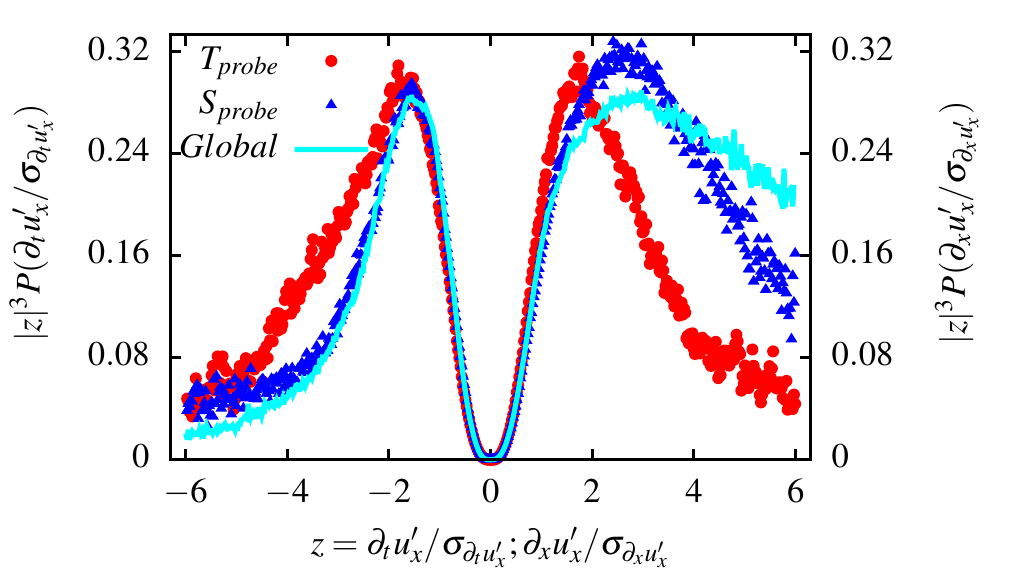}}

\subfloat[]{
\includegraphics[width=1.1\linewidth]{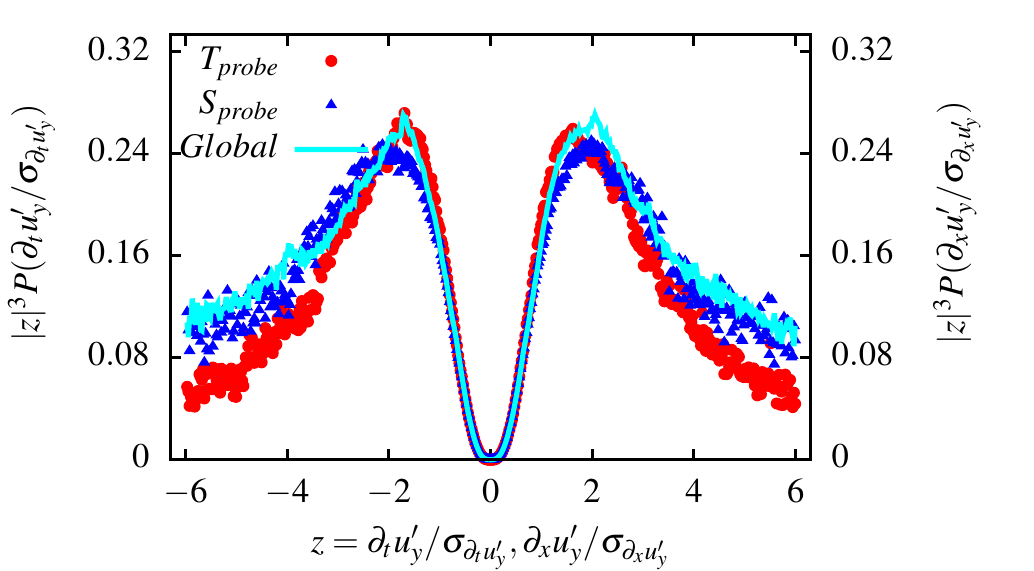}}
\caption{Third-order statistics $|z|^3P(z)$ for 
$z=\partial_t u'_i$ (local accelerations) and $z=\partial_x u'_i$ (longitudinal gradients), for $Wi=24.9$. Panel (a) refers to the streamwise velocity component ($i=x$), panel (b) to the cross-stream one ($i=y$). 
Red \textcolor{black}{(circle)} and blue \textcolor{black}{(triangle)} points respectively correspond to data from temporal and spatial probes, located where the mean flow is maximum and minimum; aqua color points \textcolor{black}{(solid line)} correspond to global spatial data. All spatial statistics are further averaged over time using $1000$ instantaneous velocity fields.
}
\label{fig: gammapdf moments of order 3}
\end{center}
\end{figure}

Further support to the differences observed above comes from an analysis of third-order statistics. The behavior of $|z|^3P(z)$, with $z$ the variable of interest and $P(z)$ the corresponding PDF
(and $\int z^3 P(z) dz$ the third-order moment of $P(z)$), is shown in Fig.~\ref{fig: pdf moments of order 3} for velocity fluctuations, and in Fig.~\ref{fig: gammapdf moments of order 3} for the temporal/spatial derivatives of the latter. 
The velocity statistics from spatial and temporal data are essentially indistinguishable (Fig.~\ref{fig: pdf moments of order 3}). Figure~\ref{fig: gammapdf moments of order 3}, instead, shows major differences between the statistics of the local accelerations $\partial_t u'_i$ and those of the corresponding spatial gradients, especially for the streamwise component $\partial_x u'_x$. A similar study on fourth, and higher, order statistics could not be carried out because of still limited convergence of the spatial data. In spite of these limitations, the above results provide, in our opinion, a clear indication of the fact that, in our flow, spatial and temporal probes sample different statistics, implying that TH cannot be considered to be satisfied.

To conclude this discussion, we note {\it a posteriori}, that such findings are consistent with the estimates of the turbulent intensities $I_{u_i} = {\sigma_{u'_i}}/{U}$, whose values, for both the $x$ and $y$ flow components, are reported in Table~ \ref{table: moments_velocity}. We find, indeed, that the typical fluctuations' intensities are not negligible with respect to the mean flow. They are close to $0.2 \, U$ in the cross-stream direction and reach $\approx 0.5 \, U$ in the streamwise one, showing that the condition in Eq.~(\ref{eq: TH1}) is not fulfilled. Important velocity fluctuations, in comparison to the mean flow, have been reported also in simulations of elastic turbulence in different geometry~\cite{canossi2020elastic}. It is worth noting, however, that the present observation shares some  similarity also with the behavior occurring in the 
3D Newtonian Kolmogorov flow at high Reynolds number~\cite{Boffetta2014}. This likely indicates that the specific values of the turbulence intensities found here may be due to the present, unbounded, geometrical configuration.

\subsection{Two-point finite-increment  statistics}

We now present the results of our analysis based on spectra and structure functions. By means of the latter, we will also address the isotropy properties of the flow at different length scales.

\subsubsection{Spectra of velocity fluctuations}

\begin{figure}[h!]
\subfloat[]{
\includegraphics[width=1\linewidth]{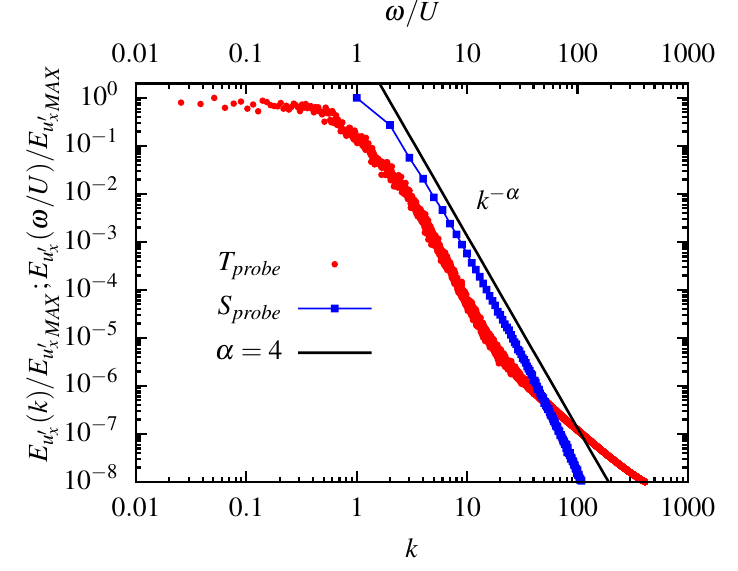}}

\subfloat[]{
\includegraphics[width=1\linewidth]{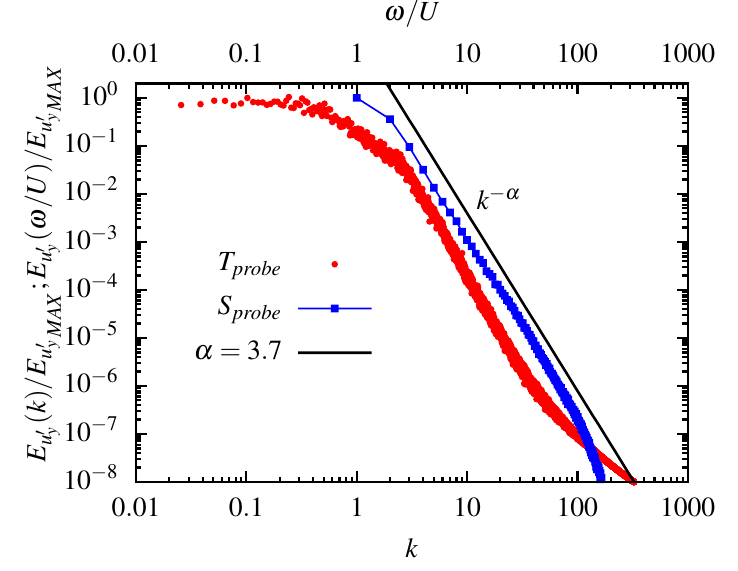}}
\caption{Spectra of velocity fluctuations $u'_x$ (a) and $u'_y$ (b) from temporal and spatial data from probes located where the mean flow is maximum and minimum, for $Wi=24.9$. The bottom horizontal axis refers to the spatial spectra, the top one refers to temporal spectra, after conversion, where $\omega=2 \pi f$, $f$ is the frequency and $U$ the mean flow intensity. All spectra in (a) and (b) are normalized by their maximum values.
}
\label{fig: unidimensional_spatial_spectra}
\end{figure}

Using  fixed-point temporal measurements, we compute 
power spectra  of velocity fluctuations, in the frequency ($f$) domain, at different probe locations. 
Aiming to compare them with unidimensional spectra, in the wavenumber ($k$) domain, computed from the corresponding spatial probes, we convert them to spatial spectra, ignoring the limited validity of TH for our flow. The resulting spectra ($E_{u'_i}$), of both type (from temporal and spatial probes), are reported in Fig.~\ref{fig: unidimensional_spatial_spectra}, with panels (a) and (b) showing the spectra of $u'_x$ and $u'_y$, respectively. 
In both cases, and independently of the considered flow component, we find power-law decaying spectra $E_{u'_i}(f) \sim f^{-\alpha}$ and $E_{u'_i}(k) \sim k^{-\alpha}$, with a common exponent $\alpha \approx 4$. The scaling range extends over almost two decades for the spatial spectra, and slightly less for the temporal ones. The non-overlapping of temporal and spatial spectra, after appropriate conversion, reflects the limited validity of TH. The similar behavior of $E_{u'_i}(f)$ and $E_{u'_i}(k)$, however, suggests that the latter may be still used as a working hypothesis, depending on what statistical features one focuses on. Finally, we remark that similarly steep spectra have been found in experiments~\cite{GS00,Groisman2004}, in the frequency domain, and theoretically predicted assuming homogeneity and isotropy~\cite{FL03}, in the wavenumber domain.  \textcolor{black}{Such a similarity is intriguing, considering that both the theory and experiments deal with 3D flows, and calls for further studies on the role of the space dimensionality in ET.}

\subsubsection{Second-order structure functions and isotropy}

We are interested in assessing the isotropy properties of the flow at different length scales. At large scales, the flow is anisotropic, as evidenced by the different rms values of velocity fluctuations in the two directions (Table~\ref{table: moments_velocity}). A natural question then is if isotropy is recovered at smaller scales. To explore this point, we consider the second-order spatial velocity structure functions:
\begin{equation}\label{Eq:structure functions X}
 S^{(2)}_{u'_i}(X) = \langle |u'_i(x+X,y,t) - u'_i(x,y,t)|^{2} \rangle_{x,y,t},
\end{equation}
with $i=x,y$, where $X$ is an increment in the streamwise direction  and $\langle ... \rangle_{x,y,t}$ indicates averaging over the spatial positions $(x,y)$ and the time $t$. The positions $(x,y)$ are taken either over all the volume in the case of \textit{global} structure functions or over lines at fixed $y$ coordinate corresponding to the positions of mean flow maxima and minima, in the so called spatial probe, $S_{probe}$, case (see Fig. \ref{Fig:1a}).
Similarly, for temporal probes, $T_{probe}$, we use a definition analogous to (\ref{Eq:structure functions X}) but for time increments at fixed ($x$) locations:
\begin{equation}
 S^{(2)}_{u'_i}(T) = \langle |u'_i(x,y,t+T) - u'_i(x,y,t)|^{2} \rangle_{y,t},
 \label{Eq:structure functions T}
\end{equation}
where the average is now over the set of $T_{probes}$ positioned at the flow maxima and minima $U_{max},U_{min}$.
The temporal increments can be converted to length increments by invoking TH, via the mean flow intensity $X=U T$.

\begin{figure}[htpb]
\subfloat[]{
\includegraphics[width=1\linewidth]{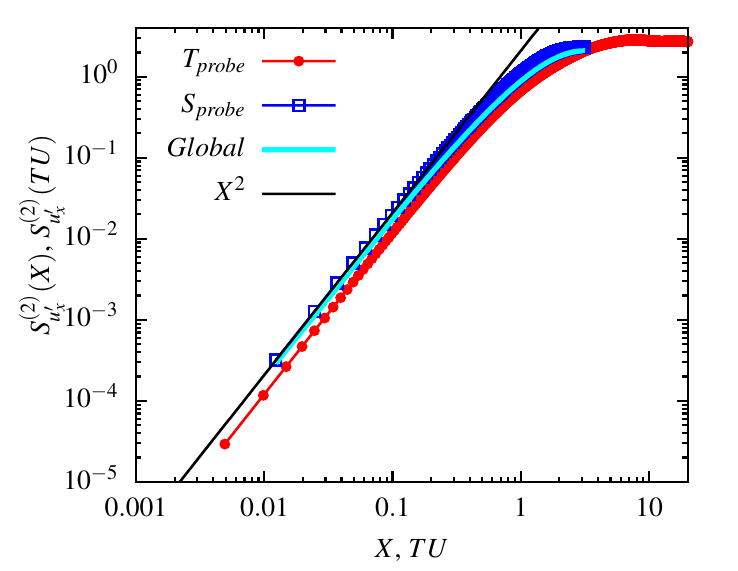}}

\subfloat[]{
\includegraphics[width=1\linewidth]{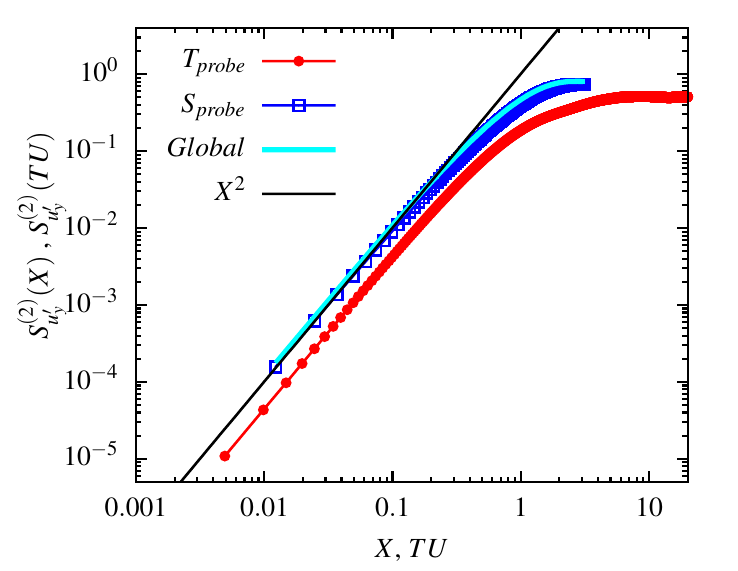}}

\caption{
Second-order velocity structure functions, for $Wi=24.9$\textcolor{black}{, versus the spatial increment (either measured directly, $X$, or via the mean flow intensity, $TU$)}. Panel (a) refers to $u'_x$, panel (b) refers to $u'_y$. 
Red \textcolor{black}{(circle)} and blue \textcolor{black}{(square)} points respectively correspond to data from temporal and spatial probes, located where the mean flow is maximum and minimum; aqua color points \textcolor{black}{(solid line)} are for global statistics. Spatial statistics are further averaged over time using $1000$ instantaneous velocity fields.
}
\label{fig: SF_order2}
\end{figure}

Let us first comment on the general behavior found for such structure functions. 
The results of our measurements are reported in Fig.~\ref{fig: SF_order2}a for $S^{(2)}_{u'_x}$ and Fig.~\ref{fig: SF_order2}b  for $S^{(2)}_{u'_y}$. From these plots, we can see that $S^{(2)}_{u'_i} \sim X^2$, meaning that the flow is smooth, which is consistent with spatial spectra being steeper than $k^{-3}$. Moreover, as for spectra, the curves from temporal and spatial probes do not overlap, even after rescaling.
Indeed, to let them superpose one would need a reduced value $\hat{U}<U=2.5$, specifically $\hat{U} \simeq 1.9$ for $u'_x$ and $\hat{U} \simeq 1.6$ for $u'_y$. This further confirms that TH does not hold in the present setup.
The tiny mismatch between the $Global$ and $S_{probe}$ functions is originated by the weak statistical non-homogeneity present in the system. This non-homogeneity however, does not affect the scaling properties.

By taking into account the statistical stationarity of the flow, the incompressibility condition and the flow isotropy, one can derive the  relation between parallel $S^{(2)}_{//}(r) = \langle (\bm{u}(\bm{x}+\bm{r},t) - \bm{u}(\bm{x},t))\cdot \bm{\hat{r}})^{2} \rangle_{x,t}$ and perpendicular $S^{(2)}_{\perp}(r) = \langle (\bm{u}(\bm{x}+\bm{r},t) - \bm{u}(\bm{x},t))\cdot \bm{\hat{n}})^{2} \rangle_{x,t}$ second-order spatial structure functions for a given increment $\bm{r} = r \bm{\hat{r}}$ and with $\bm{\hat{n}}$ the unit vector normal to $\bm{\hat{r}}$, {\it i.e.}, $\bm{\hat{n}}\cdot \bm{\hat{r}} = 0$. In a 
2D flow this reads:  
$r \frac{d}{dr} S^{(2)}_{//}(r) + S^{(2)}_{//}(r) = S^{(2)}_{\perp}(r)$ (see the complete derivation in the appendix), which  implies:
\begin{equation}
X \frac{d}{dX} S^{(2)}_{u'_x} + S^{(2)}_{u'_x} = S^{(2)}_{u'_y},
\end{equation} 
or
\begin{equation}\label{eq:iso}
\frac{d}{dX}(X S^{(2)}_{u'_x} ) = S^{(2)}_{u'_y}.
\end{equation} 

We can now observe that in the case of a smooth flow Eq.~(\ref{eq:iso}), at sufficiently small $X$, 
leads to  
$3S^{(2)}_{ u'_x} \simeq S^{(2)}_{u'_y}$. This can be also understood as the limit where $S^{(2)}_{ u'_i} = \sigma^2_{\partial_x u'_i} X^2$ and, hence, 
$S^{(2)}_{ u'_x}/S^{(2)}_{u'_y} \approx \sigma^2_{\partial_x u'_x} / \sigma^2_{\partial_x u'_y}$. 
Such a ratio of the velocity gradients' variances is then expected to attain the value $1/3$ in isotropic conditions in 2D flows, as also shown in~\cite{CalzavariniPF2020} (while it is 1/2 in 3D flows, see also the appendix). 
In the, opposite, large-scale limit ($ X \to L_0/2$ or $T \to +\infty$) the structure functions should reach a constant value. This can originate either as an artifact of the periodicity of the system (at increments $X=L_0/2$ the structure functions' slopes are 0 by construction) or by a genuine decorrelation of velocity fluctuations beyond a given scale.
In the latter case the velocity fluctuations will 
not be correlated anymore over such large increments and, consequently, from Eq.~(\ref{Eq:structure functions X}) one has $S^{(2)}_{u'_i}(X) \sim 2\langle 
{u'_i}^2 \rangle= 2 \sigma^2_{u'_i}$.
\begin{figure}[htpb]
\includegraphics[width=1.\columnwidth]{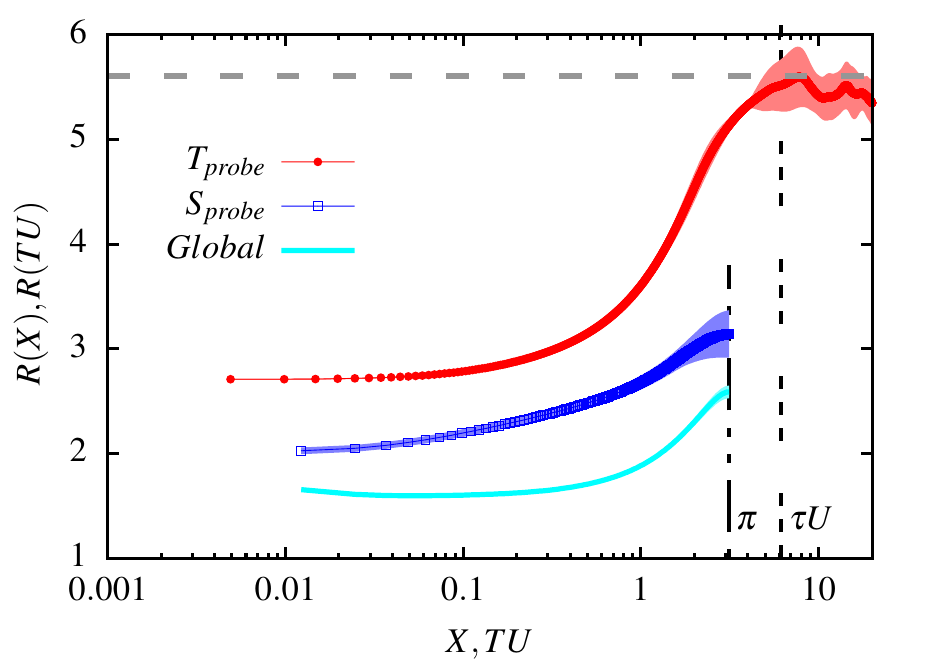}
\caption{Isotropy ratio, 
$R=S^{(2)}_{u'_x}/{S^{(2)}_{u'_y}}$, versus the spatial increment \textcolor{black}{(either measured directly, $X$, or via the mean flow intensity, $TU$)}, for $Wi=24.9$. Red \textcolor{black}{(circle)} and blue \textcolor{black}{(square)} points respectively correspond to data from temporal and spatial probes, located where the mean flow is maximum and minimum; aqua color points \textcolor{black}{(solid line)} are for global statistics. All spatial statistics are further averaged over time using 1000 instantaneous velocity fields. 
The horizontal dashed grey line corresponds to the value of $\sigma^2_{u'_x} / \sigma^2_{u'_y}$ obtained from temporal probes (Table~\ref{table: moments_velocity}). 
The vertical long-short dashed black line represents the largest possible increment, $L_0/2=\pi$. The vertical dashed black line represents the length scale corresponding to a time $t=\tau$. Error bars are estimated by dividing the data sets into two equal parts.
}
\label{fig: isotropic ratio}
\end{figure}

To test the degree of isotropy of the flow and its scale dependence, we examine the ratio 
\begin{equation}
R(X) = S^{(2)}_{u'_x}(X)/S^{(2)}_{u'_y}(X), 
\end{equation}
which is reported in Fig.~\ref{fig: isotropic ratio} as a function of the increment $X$ and again, we consider data from both spatial probes and temporal ones, computed as described above.
Our results indicate that isotropy is not recovered at small scales, as for all datasets $R(X)$ is much larger than the theoretical expectation of $1/3$. At large scales, as expected, we find anisotropic fluctuations, as the data do not approach the expected limit $\sigma^2_{u'_x} / \sigma^2_{u'_y} \approx 1$. 
The isotropy ratio from temporal probes' signals reaches 
the limit $\sigma^2_{u'_x} / \sigma^2_{u'_y} \approx 5.6$ (dashed grey curve in Fig.~\ref{fig: isotropic ratio}, obtained from the values in Table~\ref{table: moments_velocity}). The limit is attained for times larger than the polymer relaxation time $\tau$, beyond which velocity fluctuations then result to be uncorrelated, in agreement with our previous observation (see Sec.~\ref{sec:correl}). Interestingly, however, the figure also reveals that velocity increments from spatial statistics, both \textit{Global} and $S_{probe}$ ones, appear to be still correlated at the largest possible value of the increment ($X=L_0/2=\pi$), as in this case $R(X)$ does not reach the ratio of velocity variances, which should be again $\approx 5.6$ for 
$S_{probe}$ data and $\approx 4$ for 
{\it Global} ones (Table~\ref{table: moments_velocity}). 

We conclude this section by noting, once more, the lack of correspondence between spatial and temporal statistics for the flow studied here.

\subsubsection{Skewness of velocity increments}
\label{sec:4}

In this last section we want to reconsider the asymmetry observed in the PDFs of Fig.~\ref{fig: pdf_accelerations}. In particular we aim at quantifying  the normalized third-order moment of velocity increments at different scales. For this purpose we compute the scale-dependent 
quantity
\begin{equation}
    \textcolor{black}{A}_{i}(X)=S^{(3)}_{u'_i}(X)/\left[S^{(2)}_{u'_i}(X)\right]^{3/2}, \; i=x,y,  
    \label{eq:skew_scale}
\end{equation}
where $S^{(2)}_{u'_i}$ is the second-order structure function defined in Eq.~(\ref{Eq:structure functions X}) and $S^{(3)}_{u'_i}$ is the third-order structure function, defined analogously. 
The function \textcolor{black}{$A_{i}(X)$} in Eq.~(\ref{eq:skew_scale}) can be seen as a scale-by-scale skewness of velocity increments and its limit for $X \to 0$ gives the skewness of the velocity gradient \textcolor{black}{$A_{\partial_x u'_i}$}. Again, we make use of data from both spatial probes (locally and globally) and temporal ones, with time converted to space using the mean flow intensity $U$ (as in the previous section).

\begin{figure}[htpb]
\centering
\subfloat[]{
\includegraphics[width=.95\columnwidth]{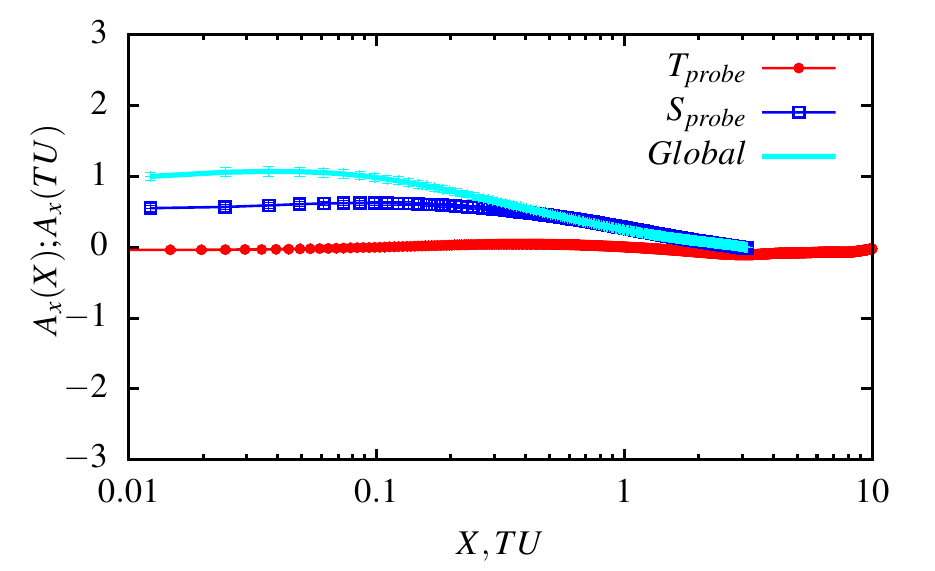}}

\subfloat[]{
\includegraphics[width=.95\columnwidth]{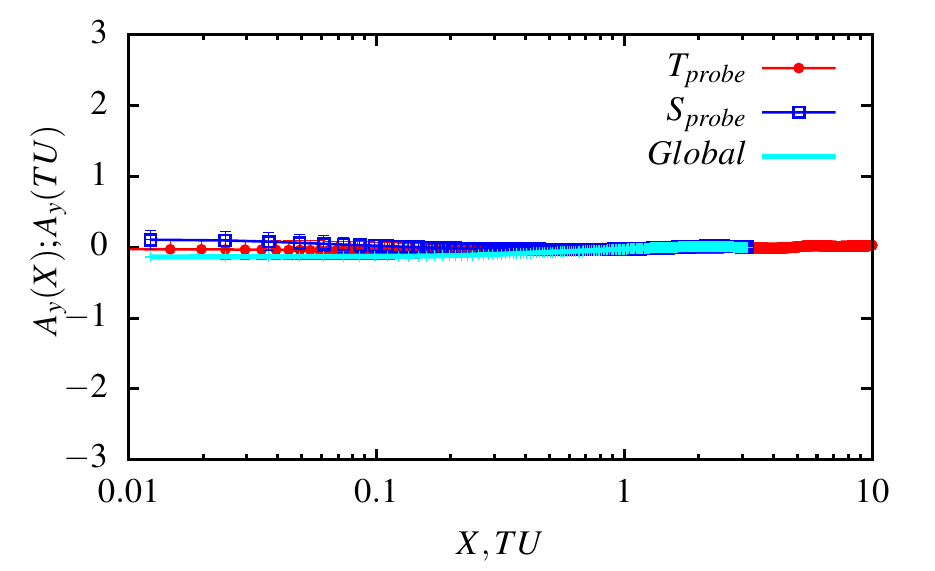}}

\caption{
Scale-dependent skewness \textcolor{black}{$A_{i}~=S^{(3)}_{u'_i}(X)/\left[S^{(2)}_{u'_i}(X)\right]^{3/2}$} versus the spatial increment \textcolor{black}{(either measured directly, $X$, or via the mean flow intensity, $TU$)}. Panel (a) refers to $u'_x$, panel (b) refers to $u'_y$. Red \textcolor{black}{(circle)} and blue \textcolor{black}{(square)} points respectively correspond to data from temporal and spatial probes, located where the mean flow is maximum and minimum; aqua color points \textcolor{black}{(solid line)} are for global statistics. All spatial statistics are further averaged over time using $1000$ instantaneous velocity fields. For all the data shown the error bar, estimated by dividing the data sets into two equal parts, is of the order of the point size.
}
\label{fig: SF_order3}
\end{figure}

The results are presented in Fig.~\ref{fig: SF_order3}, where panels (a) and (b) respectively refer to $u'_x$ and $u'_y$. While the skewness of the cross-stream-velocity increments is essentially zero at all scales (Fig.~\ref{fig: SF_order3}b), and independently of how the data are sampled, this is not always the case for the streamwise velocity. In that case in fact, only the measurement from temporal probes gives zero, scale-independent, skewness. As it can be seen from Fig.~\ref{fig: SF_order3}a, for decreasing $X$, \textcolor{black}{$A_{x}(X)$} (from spatial data) grows to attain a positive constant value, at the smallest increments, compatible with the PDFs in Fig.~\ref{fig: pdf_accelerations}a (and the value of \textcolor{black}{$A_{\partial_x u'_x}$} from Table~\ref{table: moments_spatial_gradients_global}). 
We shall stress, here, that no skewness is expected for the distribution of $\partial_x u'_x$ in inertial high-Reynolds isotropic 2D flows, as a consequence of Karman-Howarth equation~\cite{2004Davidson}. We do not have, at present, a clear explanation for the origin of the skewness in our data.
Nevertheless, we point out that these findings are in overall agreement with the analyses previously presented and indicate, through the different behaviors detected when using spatial or temporal data, that TH cannot be considered to be valid, over a whole range of scales. This, in turn, implies that temporal statistics are not apt to capture the fine statistical features of the flow, as those quantified by the third-order moment (and most likely by higher-order ones).

\section{Conclusions}
\label{sec:5} 

We have presented a systematic numerical investigation of the statistical properties of a smooth random flow of a dilute polymer solution 
resulting from the dynamics of the 2D viscoelastic Kolmogorov flow in the regime of elastic turbulence~\cite{Berti2008}.

By means of accurate temporal and spatial measurements, we find nearly Gaussian statistics of velocity fluctuations, and essentially non-Gaussian distributions, indicative of intermittency, of velocity gradients. 
These results are in overall qualitative agreement with experimental ones~\cite{Groisman2004,SACB17}, and return a picture resembling the phenomenology occurring in inertial turbulence at high Reynolds numbers~\cite{1995turbulence}.

The sampled statistics are, to good extent, homogeneous and only weakly depend on the location of the measurement. We provide evidence, however, that in our system the streamwise and cross-stream components of the velocity fluctuations do not behave equally. Indeed, our results indicate that the flow is anistropic from the largest scales, where the external forcing acts, down to the smallest ones, and for very small time lags.

The central conclusion of our study is that,  
in spite of the presence of a well-defined mean flow, in this system Taylor's frozen-field hypothesis does not fully apply. On one side, we find important velocity fluctuations, as compared to the mean-flow intensity, a feature in common with the 3D Newtonian turbulent Kolmogorov flow~\cite{Boffetta2014}. On the other side, we systematically find differences in the statistical properties computed by sampling the data in time or in space.

This fact does not reveal dramatic, as long as the focus is on the scaling properties of the spectra of velocity fluctuations, which are essentially identical in the frequency and wavenumber domains. However, as shown by our detailed investigation, statistics sampled in time do not seem to allow capturing finer features here, as those quantified by higher-order moments like the skewness of velocity increments. The present results then suggest that caution should be used when examining intermittency properties based on information from temporal sampling only.

We hope that the analysis reported in this work may help to establish sound methodologies to compare in detail numerical simulations of elastic turbulence with the available experimental results~\cite{GS00,Burghelea2007,SACB17,Steinberg2021}. This type of fine-scale comparisons appear to us crucial to improve the theoretical understanding on the subject and to validate the related numerical developments.

\appendix{}
\section{Statistical isotropy in two dimensions and second-order structure functions}
The correlation tensor of velocity fluctuations, $\bm{u}'$ in its not-normalized form is defined as:
\begin{equation}
R_{ij}(\bm{r}) = \langle u'_{i} (\bm{x}+\bm{r},t) u'_{j}(\bm{x},t) \rangle,   
\end{equation}
where $\bm{r}$ is a generic increment of magnitude $r=||\bm{r}||$ and orientation $\hat{\bm{r}} = \bm{r}/r$, and the average is taken over space and time in statistically stationary conditions (and thus the time dependence in $R_{ij}$ is dropped). 

The assumption of statistical isotropy for this second-order tensorial function can be written as:
\begin{equation}\label{eq:isoR}
R_{ij}(r) = \textcolor{black}{\mathcal{A}}(r) \hat{r}_i \hat{r}_j + \textcolor{black}{\mathcal{B}}(r) \delta_{ij}   
\end{equation}
Following \citet{1953Batchelor} we define the longitudinal (or parallel) and the 
transversal (or perpendicular) correlation functions
\begin{eqnarray}
R_{//}(r) &=& \langle (\bm{u}'(\bm{x}+\bm{r},t)\cdot \hat{\bm{r}})(\bm{u}'(\bm{x},t)\cdot \hat{\bm{r}}) \rangle\\
&=& R_{ij}\hat{r}_i \hat{r}_j =\textcolor{black}{\mathcal{A}}(r)+\textcolor{black}{\mathcal{B}}(r)= f(r) \label{eq:f(r)}
\end{eqnarray}
and 
\begin{eqnarray}
R_{\perp}(r)&=& \langle (\bm{u}'(\bm{x}+\bm{r},t)\cdot \hat{\bm{n}})(\bm{u}'(\bm{x},t)\cdot \hat{\bm{n}}) \rangle\\
&=&\textcolor{black}{\mathcal{B}}(r)= g(r) \label{eq:g(r)},
\end{eqnarray}
where $\hat{\bm{n}}$ is a unit vector perpendicular to $\hat{\bm{r}}$.
Using~(\ref{eq:f(r)}) and~(\ref{eq:g(r)}), Eq.~(\ref{eq:isoR}) can be rewritten as
\begin{equation}\label{eq:Rpp}
R_{ij}(r) = (f-g) \hat{r}_i \hat{r}_j + g(r) \delta_{ij}.
\end{equation}
Further, we note that the continuity condition $\partial_i u'_i = 0$ implies that $\frac{\partial}{\partial r_i} R_{ij}(r) = 0$ \cite{1953Batchelor}.
By using the chain rule of derivation, from (\ref{eq:Rpp}) we reach:
\begin{equation}\label{eq:chain2d}
\frac{\partial}{\partial r_i} R_{ij}(r) = 0 = \hat{r}_j \left( \frac{d}{dr}f + \frac{(f-g)}{r}\right).
\end{equation}
Since the above relation has to be true for any $\hat{r}_j$, this implies that the term between parentheses should be always zero, and so:
\begin{equation}
r \frac{df}{dr} + f = g,
\end{equation}
which is consistent with Davidson's developments (see~\cite{2004Davidson}, p.~597).\\
Finally, considering the relation between the correlation functions and the structure functions:
\begin{equation}
S_{//}^{(2)}(r)=2\left(\sigma_{u'_{//}}^2-f(r)\right), \; S_{\perp}^{(2)}(r)=2\left(\sigma_{u'_{\perp}}^2-g(r)\right),
\end{equation}
where $\sigma_{u'_{//}}^2=\sigma_{u'_{\perp}}^2$ are the variances of the longitudinal and transversal velocity fluctuations (equal under isotropy),
we obtain the notable result:
\begin{equation}
r \frac{d}{dr} S^{(2)}_{//}(r) + S^{(2)}_{//}(r) = S^{(2)}_{\perp}(r). 
\end{equation}
We shall observe that the derivation of Eq.~(\ref{eq:chain2d}) assumes a 2D space, where $\delta_{ii} = 2$. As a consequence, this result is different from the corresponding 3D expression, which takes the more familiar form \cite{1953Batchelor}: 
$$
\frac{\partial}{\partial r_i} R_{ij}(r) = 0 = \hat{r}_j \left(  \frac{d}{dr}f + 2 \frac{(f-g)}{r}\right),
$$
implying 
$$\frac{r}{2} \frac{d}{dr} S^{(2)}_{//}(r) + S^{(2)}_{//}(r) = S^{(2)}_{\perp}(r). 
$$


%

\end{document}